%% file: main.tex
\documentclass[10pt,journal,compsoc]{IEEEtran}
\usepackage{algorithm}
\usepackage{algorithmic}
\usepackage{graphicx}
\usepackage{float}
\usepackage{subcaption}
\usepackage{cancel}
\usepackage[pagebackref=true,breaklinks=true,colorlinks,bookmarks=false]{hyperref}
\input{math_commands.tex}

\def\vx{{\bm{x}}}
\def\vy{{\bm{y}}}
\def\vz{{\bm{z}}}
\usepackage[dvipsnames]{xcolor}

\ifCLASSOPTIONcompsoc
  \usepackage[nocompress]{cite}
\else
  \usepackage{cite}
\fi
\ifCLASSINFOpdf
\else
\fi
\begin{document}
%
\title{Unified Signal Compression Using a GAN with Iterative Latent Representation Optimization}
%
%
%
%

\author{Bowen~Liu,
        Changwoo~Lee, 
        Ang~Cao,  and~Hun-Seok~Kim
\IEEEcompsocitemizethanks{\IEEEcompsocthanksitem B. Liu, C. Lee, A. Cao, and H.S. Kim are with the Department of Electrical and Computer Engineering, University of Michigan, Ann Arbor, MI, 48109.\\
E-mail: \{bowenliu, cwoolee, ancao, hunseok\}@umich.edu \\

\protect}}

%
%

\markboth{}%
{Shell \MakeLowercase{\textit{et al.}}: Bare Demo of IEEEtran.cls for Computer Society Journals}
%




\IEEEtitleabstractindextext{%
\begin{abstract}
We propose a unified signal compression framework that uses a generative adversarial network (GAN) to compress heterogeneous signals. The compressed signal is represented as a latent vector and fed into a generator network that is trained to produce high quality realistic signals that minimize a target objective function. To efficiently quantize the compressed signal, non-uniformly quantized optimal latent vectors are identified by iterative back-propagation with alternating  direction  method  of  multipliers (ADMM) optimization performed for each iteration. 
The performance of the proposed signal compression method is assessed using multiple metrics including PSNR and MS-SSIM for image compression and also PESR, Kaldi, LSTM, and MLP performance for speech compression. 
Test results show that the proposed work outperforms recent state-of-the-art hand-crafted and deep learning-based signal compression methods.
\end{abstract}

}

\maketitle

\IEEEdisplaynontitleabstractindextext

%
\IEEEpeerreviewmaketitle

\IEEEraisesectionheading{\section{Introduction}\label{sec:intro}}

%
%
%
%
\IEEEPARstart{I}{mage} resolution and audio quality are continually increasing in today's data streams; however, limited communication bandwidth and storage space has posed substantial challenges to practical applications. For example, 3000 color images of 1800$\times$2400-pixels stored as raw data require 32 GB storage space, and 60 minutes stereo audio will occupy 1 GB. It is impractical to store and communicate the large data streams generated in this information-rich world, and therefore signal compression is essential. In a typical image, lower spatial frequency components carry more information than higher spatial frequency ones. This property is exploited in a number of image compression algorithms, such as JPEG and BPG. For audio information, humans can hear frequencies ranging from 20 Hz to 20 kHz but are most sensitive to sounds between 1 kHz and 5 kHz while the speech signals have high correlation in temporal domain allowing prediction of future signal. These properties are used by a number of algorithms for speech compression such as CELP and AMR.

A typical signal compression framework consists of three core components: an encoder, decoder, and quantizer. The encoder maps the target signals to the latent space, and the decoder reverses these latent representations back into target signals. The quantizer maps the signal representations to a stream of discretized symbols to reduce the bitrate of the compressed signals.

\begin{figure*}
\begin{center}
\includegraphics[width=0.9\textwidth]{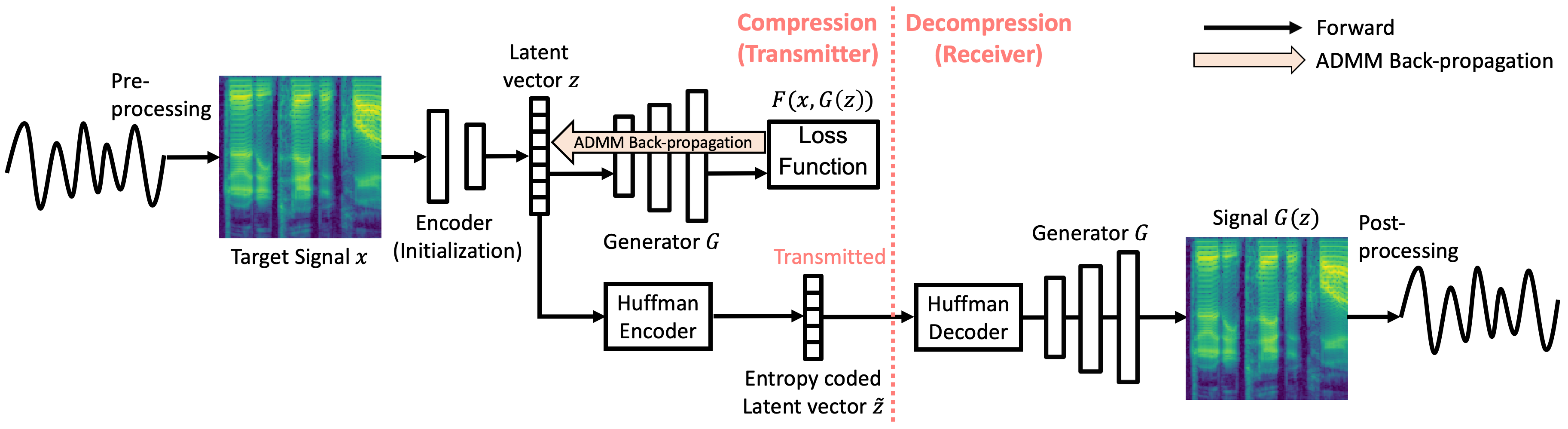}
\caption{Overview of the unified compression framework. 
The target signal is first encoded as an initialization to the latent representation by an encoder. Then the latent vector $z$ is optimized according to a reconstruction criterion through interative ADMM back-propagation (BP). The latent variable is also discretized in this iterative back-propagation process. The quantized latent signal is entropy encoded for further code size reduction. In the proposed framework, the generator parameters are shared between the transmitter and receiver. At the receiver, the generator decodes the latent signal, which is converted back to its original format via post-processing.}
\label{Frameworkoverview}
\end{center}
\end{figure*}

Inspired by the recent remarkable success of generative adversarial networks (GAN) \cite{goodfellow2014generative} in various applications, we propose a unified signal compression framework named as back-propagated GAN (BPGAN). The compressed signal is represented by a latent vector fed into a generator network that is trained to produce high-quality realistic signals (either image or speech). The core idea of BPGAN is to `search' for the optimal latent vector through iterative back-propagation during the encoding process for a given generator (with fixed weights) and compress the target signal. This process minimizes a loss function computed based on the generator output and the target signal, enabling a high-quality compressed signal represented by the latent vector input to the generator. This framework is generally applicable to different type of signals including speech and image as long as a GAN is trainable in that signal domain.

Recently, deep autoencoder structures have provided promising results in signal compression tasks. This method has achieved better performance compared with many traditional (non-learning based) compression techniques. One of the key challenges of DNN-based signal compression is optimizing the bitrate of latent representation in the auto-encoder fashion, which requires a careful design of the quantizer. Such designs commonly employ either uniform or non-uniform quantization. Deep compression\cite{han2015deep} first proposed the k-means non-uniform quantization in the DNN weight compression task, by updating the centroid of each cluster during the retraining process, to dramatically reduce the bit length of fixed point DNN weights. However, the cluster centroid values in this method vary for different compression targets, and the dynamic cluster centroid values are required during the decompression process, which adds a non-trivial overhead to the communication bandwidth. Extreme image compression \cite{leng2018extremely} proposed a differentiable quantization module inserted in the bottleneck of an encoder-decoder style network. This method optimizes the quantization center and minimizes the bitrate in an end-to-end style during the training process. In our paper, we formulate the quantization task as a manifold optimization problem with quantization constraints and put forward an alternating direction method of multipliers (ADMM) solution.

This paper is organized as follows: in Section \ref{framework} we introduce the general idea of the proposed BPGAN compression framework, including the training, compression and decompression stages. In Section \ref{implementation_details} we present the implementation details together with the data processing techniques we adopt. In Section \ref{experiments} we demonstrate and discuss the performance of our proposed method according to the designed experiments. Section \ref{conclusion} concludes the paper. 


\section{Related Work}
\label{relatedwork}
\subsection{Image Compression}
Pioneering deep autoencoder and neural network based image compression methods include \cite{balle2016end} and \cite{rippel2017real}. The focus of \cite{balle2016end} is to optimize the mean squared error (MSE) and multi-scale structural similarity for image quality assessment (MS-SSIM) between decompressed images and the originals. In \cite{theis2017lossy}, the images are compressed through an encoder, and a traditional quantization method is applied to reduce the bitrate. GAN-based models are widely explored in image compression tasks. Prior works such as \cite{mentzer2018conditional, agustsson2018generative}  apply adversarial training on deep autoencoder networks to learn the underlaying distribution of images and achieve extremely high compression rates with aesthetically pleasing details on the generated image. However, those `realistic’ details generated by the decoder often distort the ‘actual’ details of the original image. Recently, there has been a notable trend of using context-adaptive entropy models for image compression \cite{NEURIPS2018_53edebc5, lee2018contextadaptive, Cheng_2020_CVPR}, where hyperpriors allocate additional bits to more complex and bit-consuming contexts, whereas autogressive models are applied to the contexts that can be easily inferred. Unlike aforementioned prior works, our method uses a generator network for image/speech compression as well as decompression by iteratively searching the optimum latent input representation for the generator during the encoding process. 

\subsection{Speech Compression}
Traditional speech codecs such as CELP\cite{schroeder1985code}, Opus\cite{valin2012definition}, and adaptive multirate wideband (AMR-WB)\cite{bessette2002adaptive} commonly employ hand-engineered encoder-decoder datapaths. These datapaths heavily rely on crafted audio representation features and therefore require high bitrates (typically higher than 16 kbps) for acceptable speech quality. Recent DNN-based approaches including\cite{kankanahalli2018end} have demonstrated the feasibility of training an end-to-end speech codec that exhibits performance comparable to a hand-crafted AMR-WB codec at 9--24 kbps. In \cite{cernak2016composition}, paired phonological analyzers and synthesizers using deep and spiking neural networks reports a 369 bps bitrate speech codec by only keeping the content and speaker identifier information to achieve such a low bitrate.

Another important strategy to realize a high quality speech codec is to use a fine-tuned DNN-based vocoder such as Wavenet\cite{oord2016wavenet} or WaveRNN\cite{kalchbrenner2018efficient} to synthesis speech with high quality. For example, a prior work \cite{kleijn2018wavenet} employs a learned Wavenet as an encoder to generate audio given a traditional parametric codec with audio quality that is on par with that produced by AMR-WB at 23.05 kbps. Furthermore, \cite{van2017neural} demonstrates that Wavenet vocoder can generate satisfactory speech given discrete latent representation of audio generated by the vector-quantized variational autoencoder (VQ-VAE) framework. And \cite{garbacea2019low} extends this work to reach 1.6 kbps. These methods, however, do not scale well to a very low bit-rate (e.g., 1 kbps). Our BPGAN for speech compression overcomes the limitations of aforementioned prior works by achieving lower bitrates for the same quality while providing excellent scalability to trade-off bitrate vs. signal quality. 

The application of GAN has been studied for speech processing. In \cite{pascual2017segan}, speech enhancement GAN (SEGAN) is proposed, which demonstrates that GAN can achieve promising results in speech processing tasks. Furthermore, \cite{donahue2018adversarial},\cite{marafioti2019adversarial}, \cite{engel2019gansynth} show that it is possible for GAN to synthesis instrumental audio signals and simple speech with limited contexts. Nonetheless, it is still considerably difficult to generate high-quality speech audios with arbitrary latent input signals.

\subsection{ADMM Optimization}
The idea of the ADMM was first introduced in the 1970s to solve convex optimization problems with additional constraints. Later work implies that ADMM can also be used for non-convex problems and can achieve sufficiently good, if not globally optimal, solutions. The principle of the ADMM method is  splitting difficult problems into several sub-problems and solving these sub-problems efficiently by updating variables alternatively. Therefore, ADMM has been widely used in distributed optimization problems and neural network pruning tasks \cite{boyd2011distributed},\cite{leng2018extremely},\cite{zhang2018systematic}, achieving remarkable results.

\section{BPGAN: Unified Signal Compression Algorithm} \label{framework}

The proposed BPGAN compression framework is applicable to any signal type as long as it is possible to train a generative model that produces realistic outputs of that type. The overall flow of BPGAN compression is shown in Fig. \ref{Frameworkoverview}. The target signal $\vx$ is encoded as an initialization of the compressed signal $\vz_{0} = E(\vx)$ in the first step. Then, the latent vector $\vz$ is updated and optimized to minimize a specific objective function $F(\cdot)$ via iterative back-propagation through a pre-trained generator $G(\cdot)$. A typical objective function is the similarity measure between the target signal $\vx$ and the reconstructed signal $G(\vz)$. During the iterative back-propagation process, the optimal latent variable $\tilde\vz$ is discretized under the quantization scheme $Q(\cdot)$. Before transmitting to the receiver, the compressed signal is entropy encoded for further code size reduction. In the proposed framework, the generator parameters are pre-trained (i.e., fixed during the iterative back-propagation to find $\tilde\vz$) and shared between the transmitter and the receiver. At the receiver , the same generator decodes the latent signal $G(\tilde\vz)$, and this reconstructed image/speech is converted back to its original format via post-processing. The overall compression process is summarized in Algorithm \ref{BPcomalg2}.

Unlike other GAN-based approaches \cite{agustsson2018generative} relying on an encoder  to provide a compressed signal, compression in BPGAN is performed by iteratively searching and updating the generator input latent vector through back-propagation to minimize a target objective function at the output of the generator. In our scheme, the main purpose of the encoder is to accelerate the back-propagation processing by initializing the latent vector using the output of the encoder. This initialization technique reduces the number of iterations significantly. Selecting a proper optimization criterion and progressively updating the lower-dimensional representation through back-propagation for that criterion is the key step to significantly improve the quality and/or compression ratio of the signal. As this compression framework allows the application of various optimization objectives for latent representation searching, it enables objective-aware signal compression to obtain the optimized compression results tailored for a target application such as signal classification and recognition. 

\begin{algorithm}
\caption{BPGAN Compression Algorithm}
\begin{algorithmic}[1]
\REQUIRE well trained generator $G(\cdot)$, encoder $E(\cdot)$ \\ pre-defined quantization function $Q(\cdot)$
\\ signal to be compressed $\vx$
\\ objective function $F(\cdot)$
\\ quantized set $S$

\ENSURE latent vector quantization $\tilde\vz$
\STATE latent vector initialization $\vz_{0} = E(\vx)$
\STATE quantize latent elements into the discrete space $S$ \\ $\vz_{1}= Q(\vz_{0}), \vz_{1} \in \mS$
\REPEAT
\STATE calculate the objective function: $F(\vx,G(\vz_{k}))$
\STATE gradient descent: $\vz_{k+1} = \vz_{k} - \alpha \cdot \nabla F(\vz_{k})$
\STATE quantize latent elements into the discrete space $S$ \\ $\vz_{k+1}= Q(\vz_{k+1}), \vz_{k+1} \in \mS$
\UNTIL{convergence to optimal latent variable $\tilde\vz$ or maximum iteration number satisfied}
\STATE apply Huffman Coding to $\tilde\vz$
\end{algorithmic}
\label{BPcomalg2}
\end{algorithm}

\subsection{BPGAN Training Methodology}

We perform BPGAN training in a two-stage manner: pre-compression training and post-compression fine-tuning. The detailed process and objectives are discussed as follows.

\textbf{Stage one}: Train a GAN with unquantized (floating-point) latent vectors for the target signal type. This step is similar to a typical GAN training procedure \cite{wang2018high}, where a generator $G$ and discriminator $D$ are adversarially trained. In addition, an encoder $E$ is cascaded by the generator to form an auto-encoder structure, where the encoder is trained to learn mappings from the signal space to a latent space. The encoder and generator are optimized end-to-end under the following cost function:
\begin{equation}
    \min_{E,G}\max_{D}\E[D(\vx)] - \E[D(G(E(\vx)))] + \lambda_1 \cdot \E[d(\vx,G(E(\vx)))],
\end{equation}
where $G(\cdot)$, $E(\cdot)$, $D(\cdot)$ refer to the generator, encoder and discriminator respectively, and $d(\cdot)$ is the similarity measure between the original and reconstructed signals. Note that the similarity measure is different for image and speech compression and we use $d_I$ and $d_S$ to distinguish them in the following section respectively. The loss function is thus a weighted combination of the original GAN loss and the reconstruction loss.

For image compression, the similarity measure $d_I$ is defined as a weighted combination of mean squared error (MSE) and MS-SSIM loss \cite{zhao2016loss}: 
\begin{equation}
    d_{I}(\vx,G(E(\vx))) = \mathcal{L}^{\operatorname{MS-SSIM}}(\vx,G(\vz))
  +\gamma \cdot \operatorname{MSE}(\vx, G(\vz)).
\end{equation}
And the $\mathcal{L}^{\operatorname{MS-SSIM}}(\vx,\vy)$ is defined as:
\begin{equation}
\begin{split}
    \mathcal{L}^{\operatorname{MS-SSIM}}(\vx,\vy)&=\frac{1}{N} \sum (1-\operatorname{MS-SSIM}(p)) \\
    &=\frac{1}{N}\sum (1 - l_M^\alpha(p)\cdot\prod_{j=1}^M {cs_j}^{\beta_j}(p)),
\end{split}
\end{equation}
where
\begin{equation}
    l_j(p) = \frac{2\mu_{\vx}\mu_{\vy}+C_{1}}{\mu^2_{\vx}+\mu^2_{\vy}+C_{1}},
\end{equation}
\begin{equation}
    {cs_j}^{\beta_j}(p) = \frac{2\sigma_{\vx\vy}+C_{2}}{\sigma^2_{\vx}+\sigma^2_{\vy}+C_{2}}.
\end{equation}
In the above expressions, $p$ denotes a pixel located in the reference patch $\vx$ and the processed patch $\vy$, $N$ represents the total number of pixels in $\vx$ ($\vy$). $l_{j}(p)$ and $cs_{j}(p)$ are terms at the scale $j\in\{1,2,...M\}$. The means ($\mu$) and standard deviations ($\sigma$) are computed with Gaussian filters centering at pixel $p$ for each scale $j$. The standard deviation $\sigma_{G}^j$ of each Gaussian filter corresponds to the level of the constructed Gaussian pyramid. Following \cite{zhao2016loss}, we set $\gamma=0.1$ and $\alpha=\beta_{j}=1$, for $j\in\{1,2,...M\}$.


For speech compression, $d(\cdot)$ is defined as the MSE loss:
\begin{equation}
    d_{S}(\vx,G(\vz)) = MSE(\vx, G(\vz)).
\end{equation}

\textbf{Stage two}: Fine-tune the trained model with quantized latent vectors. After the first stage, we perform signal compression on the training set images through iterative back-propagation and quantization. Then we retrain the generator and discriminator models only using quantized/discretized latent vectors $\tilde{\vz}$ for model refinement in the quantized latent space. This procedure is formulated as follows:
 \begin{equation}
    \min_{G}\max_{D}\E[D(\vx)] - \E[D(G(\tilde\vz))] + \lambda_2 \cdot \E[d(\vx,G(\tilde\vz))],
\end{equation}
where $\tilde\vz$ denotes the quantized latent-vector generated by the proposed compression algorithm (using the first stage training result). 

After the above two-stage training process, we freeze the network parameters for signal compression and decompression. The generator is viewed as a fixed mapping function from the abstract latent domain to the spatial/frequency domain for image/speech during the compression process for iterative latent space searching through back-propagation.

\begin{figure}[h]
    \begin{center}
    \includegraphics[width=0.485\textwidth]{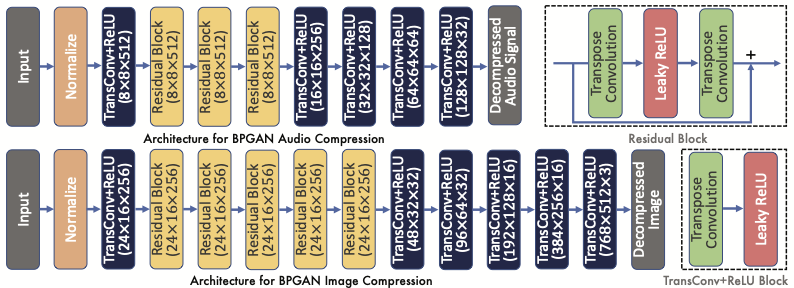}
    \end{center}
    \caption{BPGAN generator network architectures for image and speech compression with transpose convolution layers and residual blocks.}
    \label{Arch}
\end{figure}
\subsection{Objective Functions in BPGAN}
Compared with prior compression methods, a noticeable advantage of BPGAN compression lies in the fact that BPGAN can control the optimization direction of the latent representations by applying different objective functions, so that the compression can be adjusted for various application tasks without the need of retraining. From this viewpoint, the trained generator is treated as a projector mapping the latent feature space to the target signal domain.

With a well-trained generator $G$, the signal compression process is expressed as an optimization problem:
\begin{equation}
    \tilde{\vz} = \arg\min_{\vz} F(\vx,G(\vz)),
    \label{eq:opt}
\end{equation}
where $\tilde\vz$ is the optimal lower-dimensional representation of the original signal $\vx$ and $F(\cdot)$ is the objective function. Under a selected objective, we propose to solve this problem with an iterative back-propagation process based on the gradient $\nabla F(\vz)$ where $\vz$ is initialized by the encoder and regularized by the reconstruction quality criterion.

In our work, similarity measures between the reconstructed signals and the target signals are used as the optimization objectives. The conventional choices of similarity metrics are typically L1 or L2 distances; however, they are not perfectly suited for the properties of some signals and are prone to poor performance in real world applications. 
For instance, in image compression tasks, minimizing MSE would often produce blurry images and depress perceptual quality. Therefore, it is important to select/design a proper objective function that considers signal properties and back-end quality metrics. 

According to our experiments, discriminator loss gives satisfactory results in the image compression task and feature loss is advantageous for the speech compression task. The detailed discussion is extended as follows.

\subsubsection{Discriminator Loss for Image Compression}
After the adversarial training procedure for the generator and discriminator, the resulting discriminator can be regarded as a data-driven quality indicator function evaluated by a deep neural network. Therefore, we propose a discriminator loss in the compression objective function, which depicts the confidence of the discriminator on the quality of the (de)compressed signal. By introducing this loss to the compression back-propagation process, we take advantage of the complex quality measures evaluated by the discriminator for better reconstruction quality (at the cost of increased computational overhead). The total loss function used in image compression can be defined as:
\begin{equation}
    F(\vx,G(\vz)) = -D(G(\vz))) + \lambda_3 \cdot d_I(\vx,G(\vz)),
\end{equation}
where the first term gives the discriminator objective, and the second term is the combination of MS-SSIM loss and MSE between the reconstructed image and the original image. 

\subsubsection{Feature Loss for Speech Compression}
Inspired by the remarkable success of feature loss in image stylization and synthesis \cite{johnson2016perceptual,wang2018high}, we adopt and design the feature loss in our framework to improve the performance in speech signal compression. The core idea of feature loss is to apply a pre-trained classification network to the original signal and compressed signal, and compute internal activations in the network to be compared. This loss is defined in terms of the dissimilarity between those internal activations, and it is shown to yield state-of-the-art performance for various tasks without the need for prior expert knowledge or added complexity.

For better human sound perceptual quality and better back-end recognition performance, we adopt the speech feature loss to represent the content difference. Similar to image feature loss, speech feature loss measures the feature map (activation) distance between generated and real speech using a speech recognition neural network. For feature loss evaluation, we employ a VGG-BLSTM \cite{liu2019adversarial} network and train it for a phoneme recognition task on the TIMIT dataset using joint CTC-attention loss\cite{kim2017joint}. For fast inference speed of this feature loss, we only extract feature maps of the convolution layers rather than the recurrent networks. The inputs for the VGG-BLSTM network are 40 dimensional log-mel spectrograms, therefore we calculate this loss in the spectral domain. The total objective function used in speech compression is defined as:
\begin{equation}
    F(\vx,G(\vz)) = \mathcal{L}^{\mathbf{feat}} (\vx,G(\vz)) + \lambda_4 \cdot MSE(\vx, G(\vz))
\end{equation}
where $\mathcal{L}^{\mathbf{feat}}$ denotes the speech feature loss. 

\begin{figure*}[t]
    \centering
    \includegraphics[width=175mm]{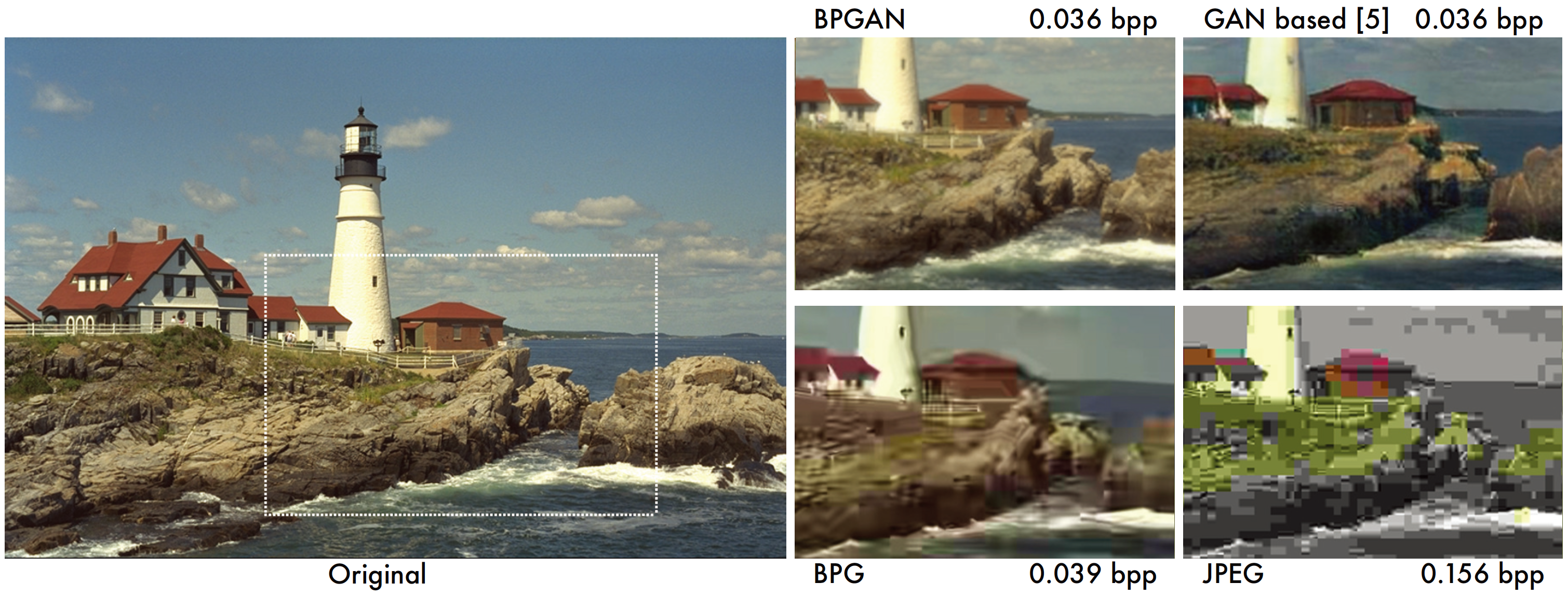}
    \caption{Compressed image from Kodak dataset. Compared with other existing methods, the lossy reconstruction of our proposed method exhibits higher fidelity with similar or lower bitrate. In the region bounded by the white box, the proposed approach exhibits the best quality in terms of preserving the color and shape of objects at a very low bit per pixel (bpp) regime.}
    \label{image_visual_1}
\end{figure*}

\subsection{Quantization in BPGAN}
To reduce the bitrate of the compressed signal, it is necessary to incorporate quantization in the compression strategy, which projects the original unquantized signal onto a discrete set of numeric values. There exists a fundamental trade-off between the number of quantization levels (bitrate) and signal quality: reducing the number of quantization levels negatively affects the compression quality while reducing the required bitrate. In BPGAN, we adopt a non-uniform quantization strategy which requires transmiting/storing the non-uniform quantization center points. Non-uniform quantization in general allows achieving higher signal quality given a fixed bitrate. However, the proper quantization centers must be chosen ahead of signal compression, which poses an additional challenge for BPGAN.

To accommodate non-uniform quantization, we propose an ADMM back-propagation algorithm, where the quantization problem is formulated as a constrained optimization task with two sub-problems. To quantify the performance of the proposed ADMM-based scheme, we compare it with alternative algorithms such as the direct quantization method and Iterative Hard Thresholding (IHT) method.

\begin{figure*}[t]
\centering
\includegraphics[width=175mm]{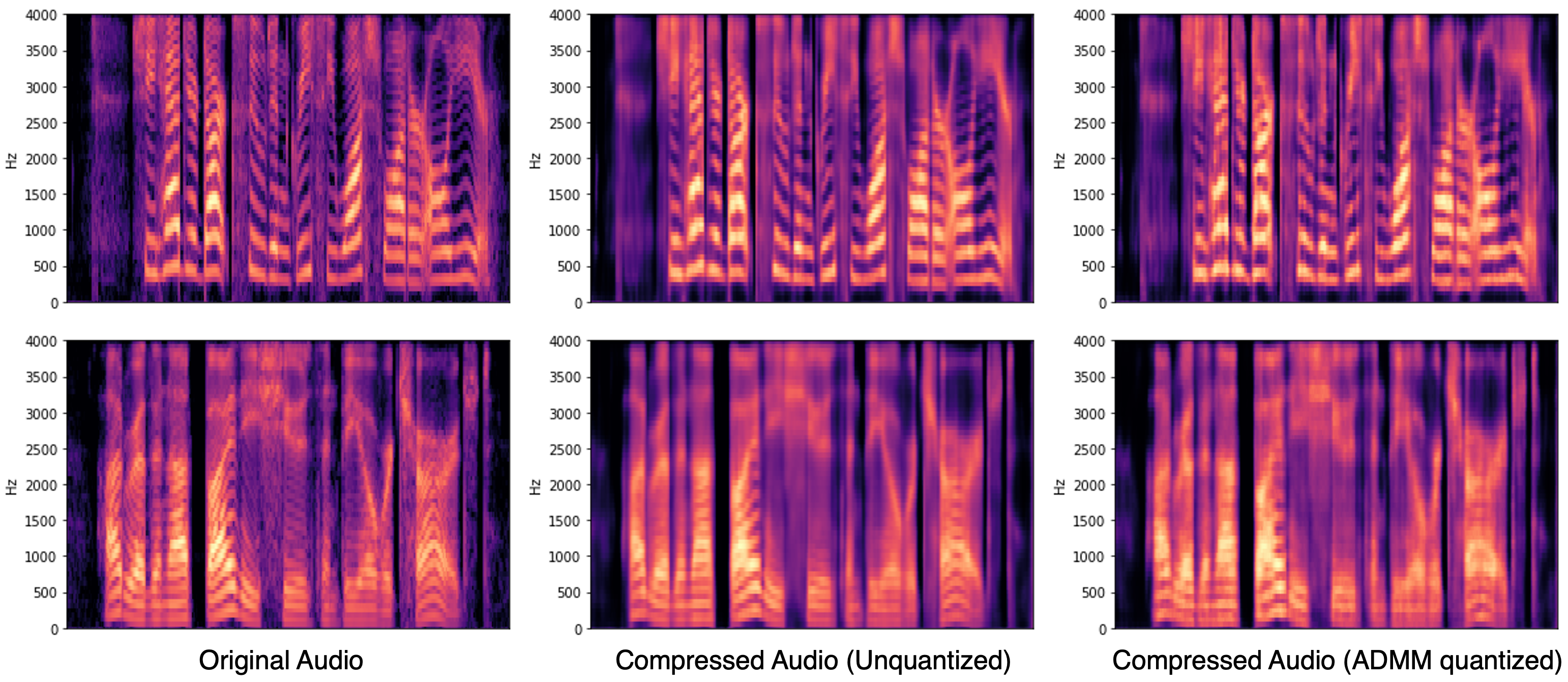}
\caption{Compressed audio on Timit dataset in the form of spectrogram. With unquantized latent vectors, the reconstructed audio (middle) preserves most of the detailed information. ADMM quantization (right) shows negligible quality degradation after quantization.}
\label{audio_visual_2}
\end{figure*}

\subsubsection{ADMM Quantization}
We formulate the latent vector search problem in the non-uniform quantized space as an optimization problem with a quantization constraint solved by ADMM \cite{ADMM}. That is, given $G(\cdot)$, $\vx$, and $F(\vx,G(\vz))$, the problem of finding an optimal quantized latent vector $\vz$ is formulated as follows:

\begin{equation}
    \arg\min_{\vz,\vu} F(\vx,G(\vz)) + I(\vu \in S) \quad s.t. {\vu=\vz}
\end{equation}
where $S$ is a non-convex set whose elements are (non-uniformly) quantized vectors, $\vu$ is an auxiliary variable, and $I(\cdot)$ is an indicator function. This optimization problem with non-convex constraints is difficult to solve directly, therefore we rewrite it as an augmented Lagrangian function and apply ADMM to solve it.

The augmented Lagrangian of the above optimization problem is given by:
\begin{multline}
    L(\vz,\vu,\bm{\eta},\mu) = F(\vx,G(\vz)) + \\ I(\vu \in S) + \frac{\mu}{2}(\|\vz - \vu + \bm{\eta}) \|^{2}_{2} - \|\bm{\eta}\|^2_2).
    \label{eq:12}
\end{multline}

ADMM is designed to minimize $ L(\vz,\vu,\bm{\eta},\mu)$ by updating variables $\vz, \vu, \bm{\eta}$ alternatively in every iteration to approach the optimal values that minimize the quantization error. The ADMM updating procedure for $k=0,1,2,\dots$ is given by:
\begin{gather}
    \vz_{k+1} =  \arg \min_{\vz} F(\vx,G(\vz)) + \frac{\mu}{2}\|\vz - \vu_{k} + \bm{\eta}_{k}\|^2_2
   \\
   \vu_{k+1} = \arg \min_{\vu} I(\vu \in S) + \frac{\mu}{2}\|\vz_{k+1} - \vu + \bm{\eta}_{k}\|^2_2
   \label{eq:14}
   \\
   \bm{\eta}_{k+1} = \bm{\eta}_{k} + \vz_{k+1} - \vu_{k+1}.
\end{gather}

The optimization criteria for the compressed signal $\vz$ described in (\ref{eq:12}) is additionally regularized with an $L_2$ term. This can be easily translated to an updating rule with an stochastic gradient descent (SGD) or Adam\cite{kingma2014adam} optimizer.

For the updating objective of $\vu$ depicted in (\ref{eq:14}), the analytical solution is $\vu_{k+1} = Q(\vz_{k+1}+\bm{\eta}_{k})$ where $Q(\cdot)$ is a non-uniform quantization function that directly projects into the set of quantized set $S$. The non-uniform quantization centers of $S$ are obtained by K-means clustering based on the distribution of the unquantized value of the latent vectors corresponding to the entire training dataset. The quantization centers are obtained during the training and unchanged for each signal reconstruction to avoid separately transmitting the quantization codebook for each target signal. To ensure that the latent $z$ is quantized after the alternating operations, we projects the latent vector to the discrete set $S$ directly as the final step of ADMM quantization.

With ADMM quantization being specified, we refine the proposed algorithm as described in Algorithm \ref{BPcomalg1}. The idea behind ADMM back-propagation is that this algorithm searches for the optimal point or local optimal point in the discretized latent space to preserve signal fidelity after reconstruction.

\begin{algorithm}
\caption{ADMM-based BPGAN Compression Algorithm}
\label{BPcomalg1}
\begin{algorithmic}[1]
\REQUIRE well trained generator $G(\cdot)$, encoder $E(\cdot)$ \\ pre-defined quantization function $Q(\cdot)$
\\ signal to be compressed $\vx$
\\ objective function $F(\cdot)$
\\ set hyperparameters $\mu, \alpha$
\ENSURE quantized latent vector $\tilde\vz$
\STATE latent vector initialization $\vz_{0} = E(\vx)$
\STATE auxiliary variables initialization: $\vu_{0}= Q(\vz_0)$, $\eta_{0}=0$
\REPEAT
\STATE calculate the optimization objective: \\$O(\vz_{k}) = F(\vx,G(\vz_{k})) + \frac{\mu}{2}\|\vz_{k} - \vu_{k} + \bm{\eta}_{k}\|^2_2$
\STATE gradient descent: $\vz_{k+1} = \vz_{k} - \alpha \cdot \nabla O(\vz)$
\STATE $\vu$ update: $\vu_{k+1} = Q(\vz_{k+1}+\bm{\eta}_{k})$
\STATE dual ascent: $\bm{\eta}_{k+1} = \bm{\eta}_{k} + \vz_{k+1} - \vu_{k+1}$
\UNTIL{convergence to $\tilde\vz$ or maximum iteration number satisfied}
\STATE apply Huffman Coding to $\tilde\vz$
\end{algorithmic}
\end{algorithm}

\subsubsection{Direct Quantization}
A baseline method to be compared with the proposed ADMM based approach is direct quantization, where an unquantized latent vector is first obtained by back-propagation and then each element is projected to the nearest numerical value in the quantized set. Then the iteration continues by updating the latent by back-propagation to a new unquantized vector. Applying this direct quantization one-time only to the final latent vector will degrade the fidelity of the reconstructed signal. Therefore, quantization and backpropagation steps are operated in an alternating order. 

\subsubsection{Iterative Hard Thresholding (IHT) Quantization}
Iterative Hard Thresholding (IHT) is a very popular technique in compressive sensing, which is designed to estimate the sparse signal $\vx$ with the problem formulation:
\begin{equation}
    \vy = \Phi  \vx +\ve
\end{equation}
where $\vy \in \R^M $ is the observation signal, $\vx \in \R^N$ is the sparse signal to be estimated, $\Phi \in \R^{M\times N}$ is the observation matrix and $\ve$ is the observation noise.

The principle optimization iteration in the IHT \cite{blumensath2009iterative} is:
\begin{equation}
    \vx_{n+1} = H_s(\vx + \Phi^{T}(\vy-\Phi\vx))
\end{equation}
where $H_s(\va)$ is the non-linear operator that sets all but the largest (in magnitude) $s$ elements of $\va$ to zero. The IHT algorithm is adopted in the neural network weight pruning tasks \cite{yuan2014gradient,jin2016training}, showing significant sparsity enhancement of the neural network weights while maintaining the accuracy.

We adopt IHT in the same manner as the Gradient Hard Thresholding Pursuit (GradHTP) used in \cite{yuan2014gradient}: the whole optimization process is divided into several sub-steps, and in each sub-step we discretize a certain number of elements.

More specifically, we divide the total iteration process into N sub-steps and in sub-step $i$, we quantize unfrozen elements with $M_i$ number and freeze them and then update the remaining unfrozen elements. The quantized elements in each sub-step are selected by the distance to the quantization center and this algorithm terminates when all of the N substeps are finished, which implies that all of the elements are fixed and quantized. This quantization process is depicted in Algorithm ~\ref{ihtalg}.

\begin{algorithm}
\caption{IHT-based BPGAN compression Algorithm}
\label{ihtalg}
\begin{algorithmic}[1]
\REQUIRE well trained generator $G$, encoder $E$ \\ pre-defined quantization function $Q(\cdot)$
\\ signal to be compressed $\vx$
\\ loss function $F(\cdot)$
\\ set sub-step number $N$, iteration number $n_i$ in sub-step $i$, quantization number $M_i$ in sub-step $i$
\ENSURE latent Vector quantization $\tilde\vz$
\STATE latent vector initialization $\vz_{0} = E(\vx)$
\FOR{sub-step i in N}
\STATE let S is the elements set of which are not frozen 
\FOR{sub-step iteration number $n_i$}
\STATE $\vz_{k+1} = \vz_{k} - \alpha \cdot \nabla F(\vz)$,  subject to $\vz_{k+1} \subseteq$ S
\ENDFOR
\STATE calculate the distance between original elements and quantized elements for elements in S
\STATE quantize $M_i$ unfrozen elements that give the least distance
\STATE update unfrozen set S
\ENDFOR
\STATE apply Huffman Coding to $\tilde\vz$
\end{algorithmic}
\end{algorithm}




\begin{figure*}[htbp]
\centering
\includegraphics[width=175mm]{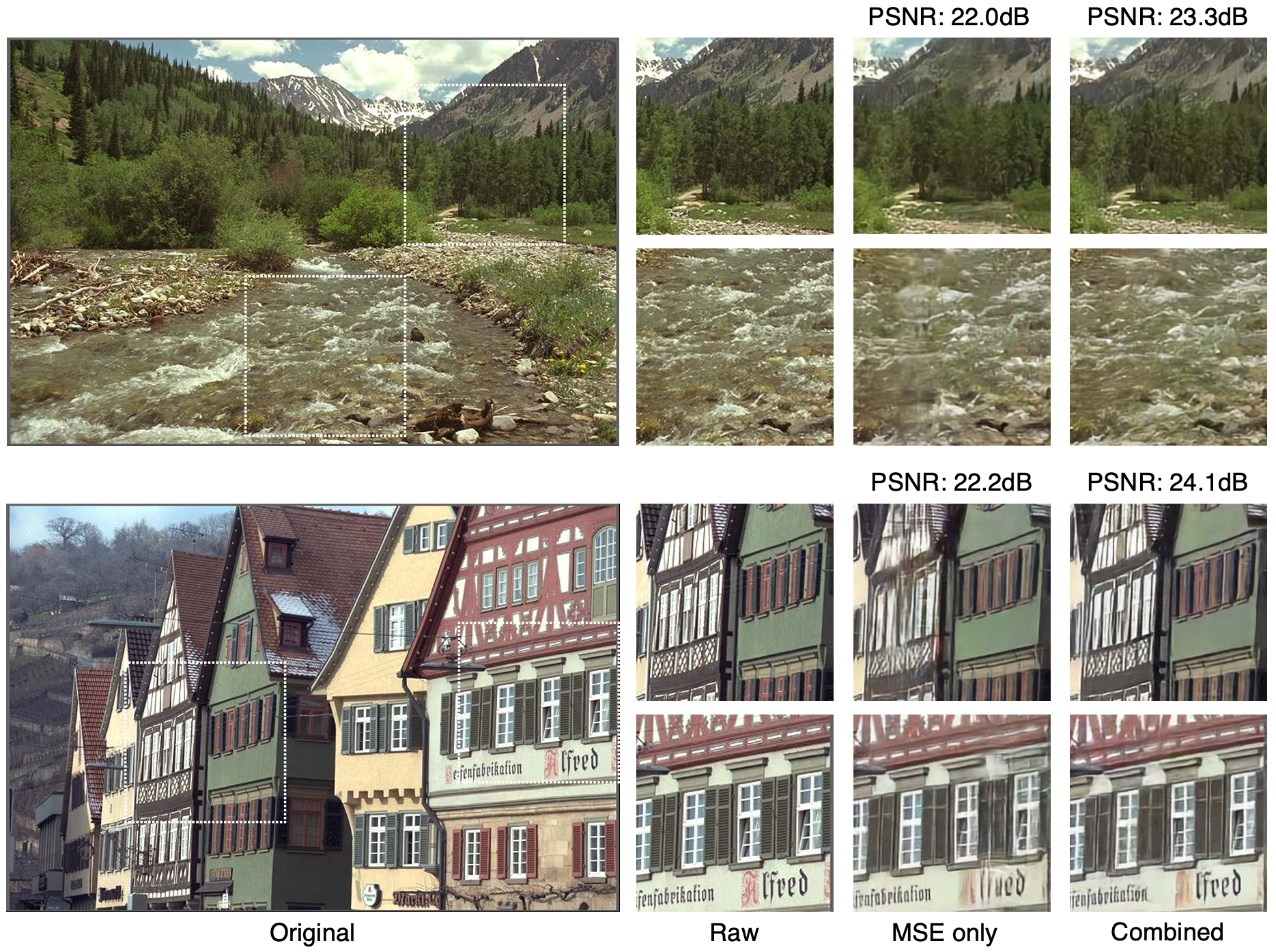}
\caption{Visualization of reconstructed images optimized according to MSE and the joint objective that combines MSE with discriminator output. Images on the right are substaintially better than the middle ones with higher PSNR and better visual quality. In both cases (top and bottom), using the joint objective facilitates sharper and better defined details in the reconstructed images. The observation indicates that incorporating perceptual objectives with the per-pixel similarity measure enables the search of latent representations that can capture more accurate features for visually pleasing reconstructions.} 
\label{image_visual_3}
\end{figure*}

\subsection{Huffman Encoding}
To reduce the code size further, we apply Huffman coding to the quantized latent vectors. The Huffman coding is a widely used prefix-free entropy coding method \cite{van1976construction} that assigns different lengths of codes to encoded symbols depending on the relative frequency of the corresponding symbols. By assigning shorter codes to symbols with frequent occurrence, Huffman coding reduces the bitrate without losing any information. 

\subsection{Latent Initialization with Trained Encoder}
Finding sufficiently good latent vectors by back-propagation with random initialization could require many iteration steps and might consume excessive computation resources. To accelerate this process, we propose to train an encoder that learns mappings from target signals to (optimal) latent vectors. The output latent vectors of the encoder often cannot be exactly the same as the target optimal vectors but are expected to approximate them so that they provide good initialization for interative back-propagation. The encoder is trained with the generator and discriminator in the training stage and it produces unquantized latent vectors to initiate the optimization process.

\section{Implementation Details} \label{implementation_details}
The BPGAN compression is a unified signal compression method for different types of signal (e.g. image and speech). To apply the unified approach, pre-processing and post-precessing are crucial for our algorithm during the training phase and the compression process. In this section, we explain in details on how to apply image and speech signals into the BPGAN framework, and also the hyperparameter settings during training. 

\subsection{Image Compression}
We adopt the Adam optimizer for training and set the batch size to 128. The network is trained for 100 epochs in total, with a fixed learning rate of 0.0005 in the first 50 epochs and a linear decayed learning rate in the following 50 epochs. To better evaluate the performance of our algorithm on different datasets, images in the training dataset are resized to a pre-defined resolution (768$\times$512) in RGB format.

\begin{table*}[ht]
\caption{Image compression performance comparison.}
\label{image_comp_perf}
\centering
\resizebox{1\textwidth}{!}{%
\begin{tabular}{c|c|c|c|c|c}
\hline
\multicolumn{1}{c}{\bf Method}  
&\multicolumn{1}{c}{\bf Bitrate (bit per pixel)}
&\multicolumn{1}{c}{\bf PSNR} 
&\multicolumn{1}{c}{\bf MS-SSIM}
&\multicolumn{1}{c}{\bf ImageNet Top-1 error rate \%}
&\multicolumn{1}{c}{\bf ImageNet Top-5 error rate \%}
\\
\hline
Original & 24 & - & - & 23.7 & 6.8 \\
\bf Ours (BPGAN) & \bf 0.286 & \bf 32.9 & \bf 0.968 & \bf 23.7 & \bf 6.8 \\
GAN based \cite{agustsson2018generative} & 0.305 & 28.2 & 0.922 & 26.0 & 7.9 \\
JPEG & 0.306 & 26.9 & 0.864 & 42.5 & 16.6 \\
BPG & 0.298 & 32.3 & 0.961 & 25.8 & 7.4 \\
\hline
 \end{tabular}
    }
\end{table*}

\subsection{Speech Compression}
\subsubsection{Training Settings}
During training, we employ the Adam optimizer and set the batch size to 128. The framework is trained for 200 epochs in total, with a fixed learning rate of 0.0002 in the first 100 epochs and a linear decayed learning rate in the following 100 epochs. 

\subsubsection{Speech Pre-processing}
For BPGAN compression, speech signals sampled in time domain are converted to spectral domain representations using short-time Fourier transform (STFT) with 128-sample stride and 512-sample frame size, resulting in 75\% frame overlap. BPGAN only uses the magnitude information of the STFT output (i.e., spectrogram) for signal compression. Our experiments show that a higher overlap ratio can lead to better audio quality with a higher bit rate, which supports the claim of \cite{marafioti2019adversarial}. To further reduce the bit rate of compressed signal, spectrograms are transformed into mel spectrograms with 128 mel frequency bins for each frame. We collect 128 frame mel frequency bins to make up one 128$\times$128 mel spectrogram, which corresponds to one second of speech audio at a sample rate of 16 kHz.

For human sound perception-oriented compression, we calculate the log-magnitude of mel-spectrograms and normalize them for compression \cite{marafioti2019adversarial}. We first normalize the magnitudes of mel-spectrograms to have a maximum value of 1, limiting the log-magnitude value to the range of $[-\infty,0]$. Then we limit the dynamic range of the log-magnitude into $[-r,0]$ by truncating by $-r$. Finally we scale and shift it back to $[-1,1]$ for the generator. In our experiments, we set $r=8$.

\subsubsection{Speech Post-processing for Phase Recovery}
Inverse STFT requires magnitude and phase information to recover the time domain speech audio signals. However, phase information is removed in our pre-processing and the following BPGAN compression. Therefore we need to estimate phase information given magnitudes, which can be regarded as an additional decoding process.  

In our work, Phase Gradient Heap Integration (PGHI) \cite{pruuvsa2017noniterative} method is employed to recover the phase from amplitude-only signal. This technique relies on magnitude-phase gradient integration relations for phase reconstruction. PGHI often significantly outperforms Griffin-Lim \cite{perraudin2013fast} and shows superior robustness because it avoids integrating through phase instability areas.

\section{Experiments} \label{experiments}



  In this section, we compare the proposed approach with conventional and learning-based image/speech compression methods in the perspective of various evaluation metrics. To qualtify the impact of various techniques incorporated in our work, we perform ablation studies on ADMM quantization, latent representation optimization objectives, and latent initialization with a trained encoder in the following subsections.
  
  The code to learn the model and experiment details is available at: \href{https://github.com/BowenL0218/BPGAN-Signal-Compression}{\textit{https://github.com/BowenL0218/BPGAN-Signal-Compression}}.

\begin{table*}[ht]
    \caption{Speech compression performance comparison.}
    \label{audio_comp_perf}
    \centering
    \resizebox{1\textwidth}{!}{%
    \begin{tabular}{c|c|c|c|c|c|c}
    \hline 
\multicolumn{1}{c}{\bf Method}  
&\multicolumn{1}{c}{\bf Bitrate (bps)}
&\multicolumn{1}{c}{\bf PESQ} 
&\multicolumn{1}{c}{\bf MUSHRA}
&\multicolumn{1}{c}{\bf Kaldi Percentage \%}
&\multicolumn{1}{c}{\bf MLP Percentage \%}
&\multicolumn{1}{c}{\bf LSTM Percentage \%}
\\
\hline
Original & 256k       & 4.50 & 95.0  &18.7 &18.6 &15.4 \\
\bf Ours (BPGAN) & \bf 2k       & \bf 3.25 & \bf 64.1  &\bf 20.9 &\bf 20.8 &\bf 18.6 \\
    CELP &4k        & 2.54 & 32.0  &28.2 &27.6 &27.3 \\
    CELP &8k        & 3.39 & 59.4  &23.0 &23.6 &21.2 \\
    Opus &9k        & 3.47 & 79.3  &22.7 &23.7 &21.2 \\
    AMR  &6.6k      & 3.36 & 58.9  &22.6 &23.6 &22.3 \\
\hline
\end{tabular}}
\end{table*}

\subsection{BPGAN Image Compression}
\subsubsection{Datasets}
\textbf{Open Images Dataset V5.}
The proposed image compression framework is trained on Open Images Dataset V5, which contains 9 million images belonging to 600 boxable categories. The images in this dataset are diverse and contain complex scenes with 8 objects per image on average. Comprising an abundant amount of various contents, it helps the generator model to learn the diversity of real world objects. As the GAN network uses the same sized images for its training input, images are rescaled to 768 $\times$ 512 pixels prior to training.

\textbf{Kodak Dataset.}
The Kodak dataset comprises 25 landscape or portrait RGB images of size $768\times 512$ pixels. It is widely used in image compression tasks and serves as a generic benchmark data collection. We use Kodak dataset to evaluate PSNR and MS-SSIM of (de)compressed images using the generator trained with Open Images Dataset V5. Kodak dataset is not used during the training of the generator, discriminator, and encoder.

\textbf{ImageNet Dataset.}
The ImageNet 2012 classification dataset consists of images belonging to 1000 classes. It is split into three sets: training (1.3M images), validation (50K images), and testing (100K images with held-out class labels). ImageNet dataset is used to quantify the image quality by performing ImageNet classification on (de)compressed images using  VGG-16 network pre-trained with raw images.

\subsubsection{Evaluation Metrics}
The quantitative performance of the proposed image compression method is primarily measured by PSNR and MS-SSIM on the Kodak dataset which is a widely used generic benchmark data collection to evaluate image compression. A significant drawback of Kodak is that the limited amount of data which may introduce biases to the quality measurement. To alleviate this issue, we also evaluate our image compression method by comparing the classification accuracy on ImageNet using the original test dataset and the dataset processed by image compression methods. For subjective image quality evaluation, we present sample reconstructed images in 
Fig. \ref{image_visual_1} and Fig. \ref{image_visual_3}. 

\subsubsection{Results and Analysis}
We compare the proposed method with conventional codecs such as BPG \cite{BPG} and JPEG, and an existing GAN-based method \cite{agustsson2018generative}. 
Notably for extremely low-bpp (bit per pixel) regimes, PSNR and MS-SSIM are usually inadequate to reflect the perceptual image fidelity. Thus we report the classification error rate metric as the VGG-16 neural network-interpreted image quality for ImageNet dataset. To evalute the classification accuracy, our generator is modified and trained to match the dimension of ImageNet images (256$\times$256 in RGB format).  

Our experimental results for image compression are shown in Fig. \ref{image_visual_1} and summarized in Table \ref{image_comp_perf}. PSNR and MS-SSIM evaluations are performed on the Kodak dataset with 768$\times$512 resolution images in RGB format (24-bit per pixel, bpp). A compressed image of 0.286bpp is obtained by using a latent vector $\vz$ of 20000 dimensions and non-uniform quantization with 256 levels for each element before Huffman coding. 

In Table \ref{image_comp_perf}, we adopt an extremely high compression ratio (84$\times$) to show the performance comparison between different methods. Results measured by the classification accuracy metric (error rate) are listed in the last two columns. Table \ref{image_comp_perf} indicates that the proposed BPGAN compression  produces higher quality images measured by PSNR and MS-SSIM with a compressed rate of $\approx 0.3$ bpp compared to other existing compression methods. Our approach manages to maintain the same classification accuracy before and after compression of the ImageNet test images. The result shows that our approach, unlike other approaches, successfully preserves the perceptual features of the target images after compression without significantly affecting deep visual learning.

The compressed images generated by our method have higher subjective quality for the same/similar bitrate compared to other methods. In Fig. \ref{image_visual_1}, we visualize the compressed image comparison of BPG, JPEG, and \cite{agustsson2018generative} with a high compression ratio (0.036 bpp) using \textit{Kodim21} in the Kodak dataset. With severe block artifacts, JPEG fails to compress this image at the target bpp of 0.036. BPG handles the high-frequency contents better (e.g., the fence in the image), however, it loses some of the features of the rocks and waves. The image generated by the method described in \cite{agustsson2018generative} provides smoother features especially in regions containing clouds. It is worth noting that a GAN based method \cite{agustsson2018generative} sometimes `creates' realistic details in the compressed image that are non-existing in the original. The actual details are better preserved in the BPGAN compression output because of the joint loss (discriminator score, MSE, and MS-SSIM) used to find the optimal compressed latent in the compression stage.

\begin{figure*}[t]
\centering
   \begin{subfigure}{.5\textwidth}
   \centering
    \includegraphics[width=\linewidth]{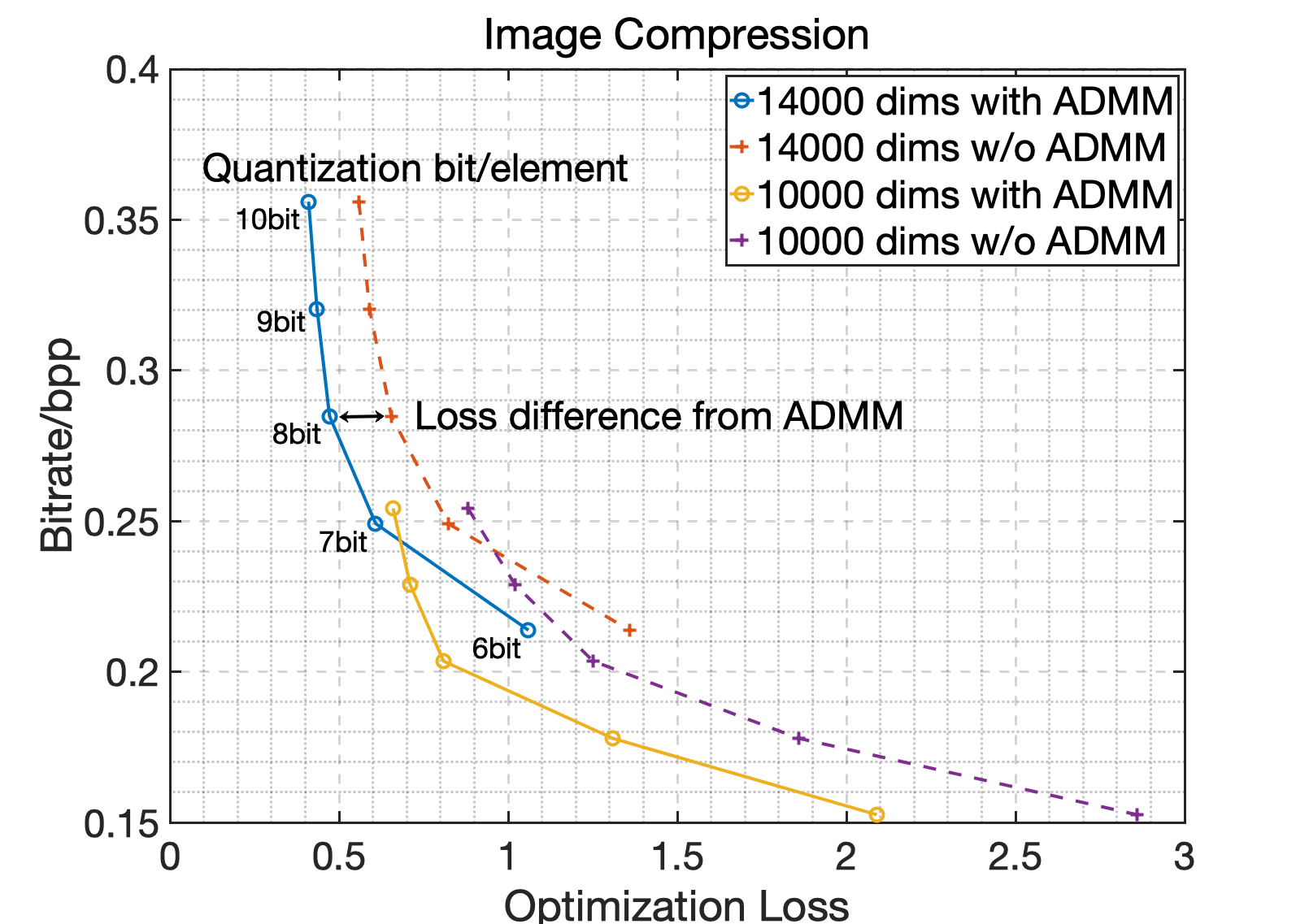}
    \end{subfigure}%
    \begin{subfigure}{.5\textwidth}
    \centering
    \includegraphics[width=\linewidth]{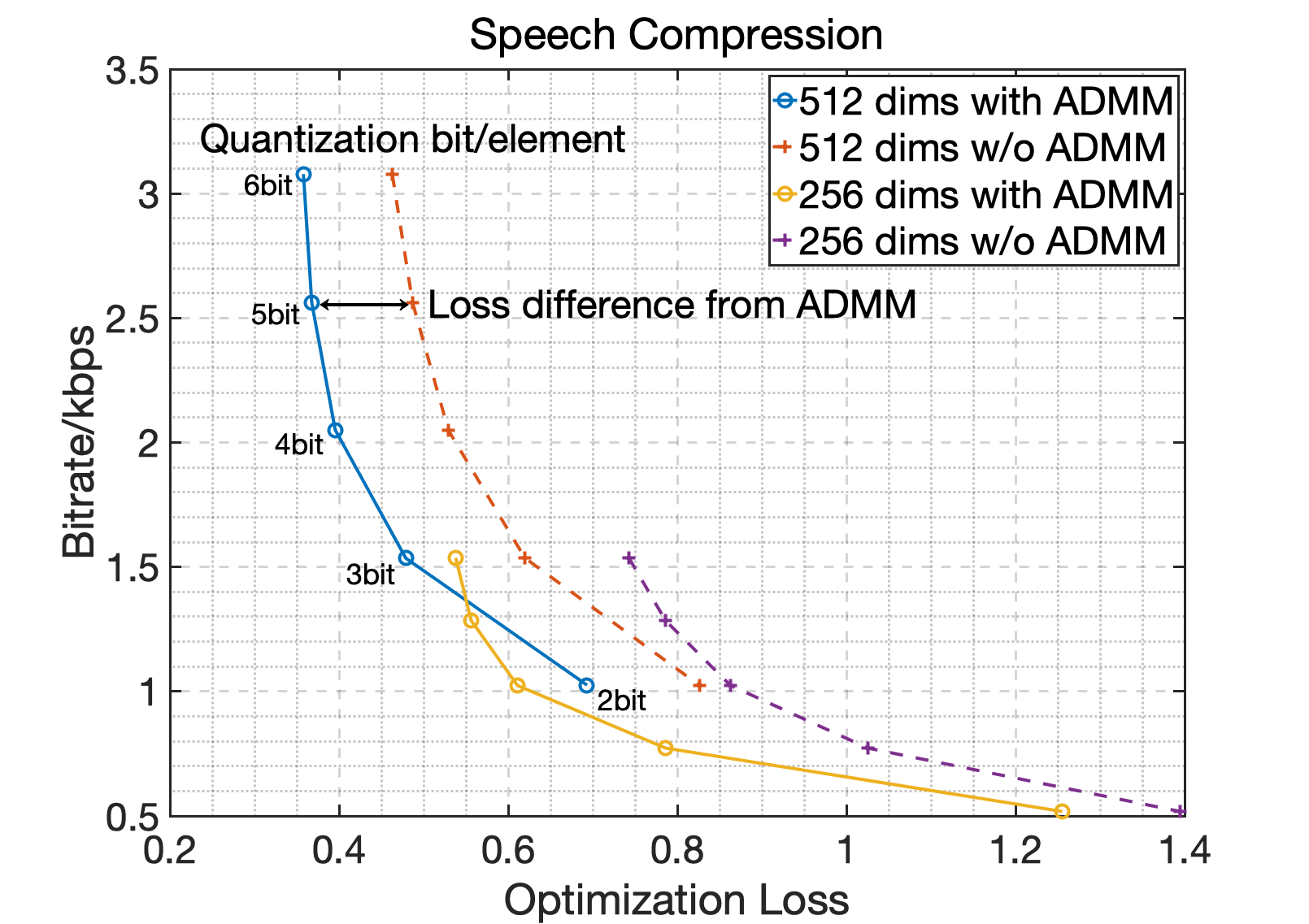}
    \end{subfigure}
\caption{Image and audio compression parameter sensitivity evaluation: rate-distortion trade-off is obtained by adjusting the latent vector size and number of quantization levels. BPGAN offers a wide rate-distortion trade-off space. Compared with direct quantization, ADMM leads to substantially lower optimization loss under the same latent dimension and quantization level.}
\label{admm_loss}
\end{figure*}

\subsection{BPGAN Speech Compression}
\subsubsection{Datasets}
Our experiments are performed on the TIMIT dataset, which contains a total of 6300 sentences spoken by 630 speakers from 8 major dialect regions of the United States at a sample rate of 16 kHz. We use a training subset for model training and then evaluate on the testing subset. The training and testing subsets are mutually exclusive.
\begin{figure}[htbp]
\centering
\includegraphics[width=0.5 \textwidth]{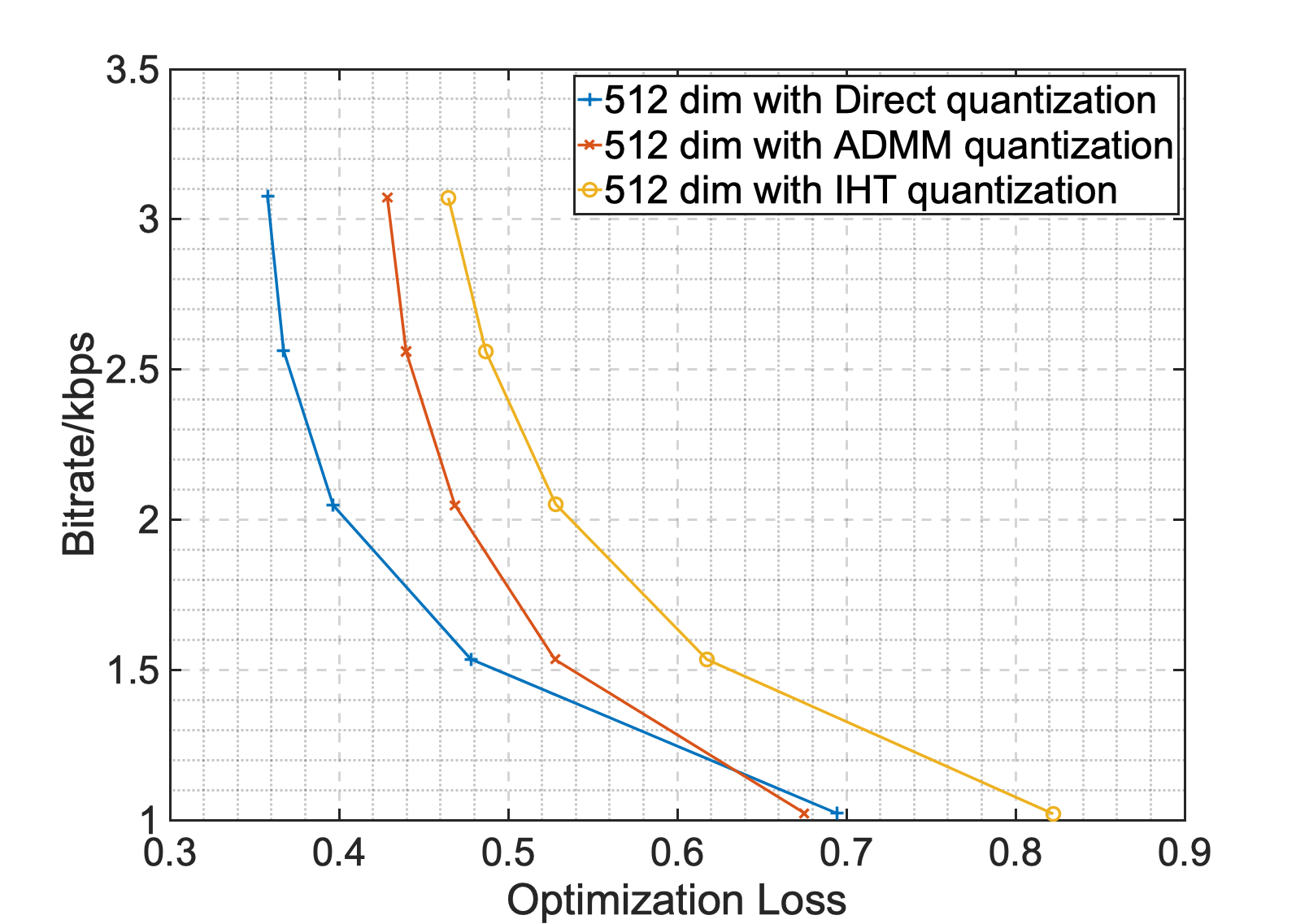}
\caption{Optimization loss vs. bitrate comparison for different quantization methods. ADMM outperforms direct quantization and IHT.}
\label{opt_comparison}
\end{figure}
The log-magnitude mel-spectrum used in training is computed with 128 mel-frequency bins per frame. And we separate the whole mel-spectrum into 128x140 sized patches (140 frames per patch) with an overlap of 12 frames. For producing the testing data, we directly separate the whole mel-spectrum into patches without overlap.

\begin{figure*}[t]
\centering
   \begin{subfigure}{.5\textwidth}
   \centering
    \includegraphics[width=\linewidth]{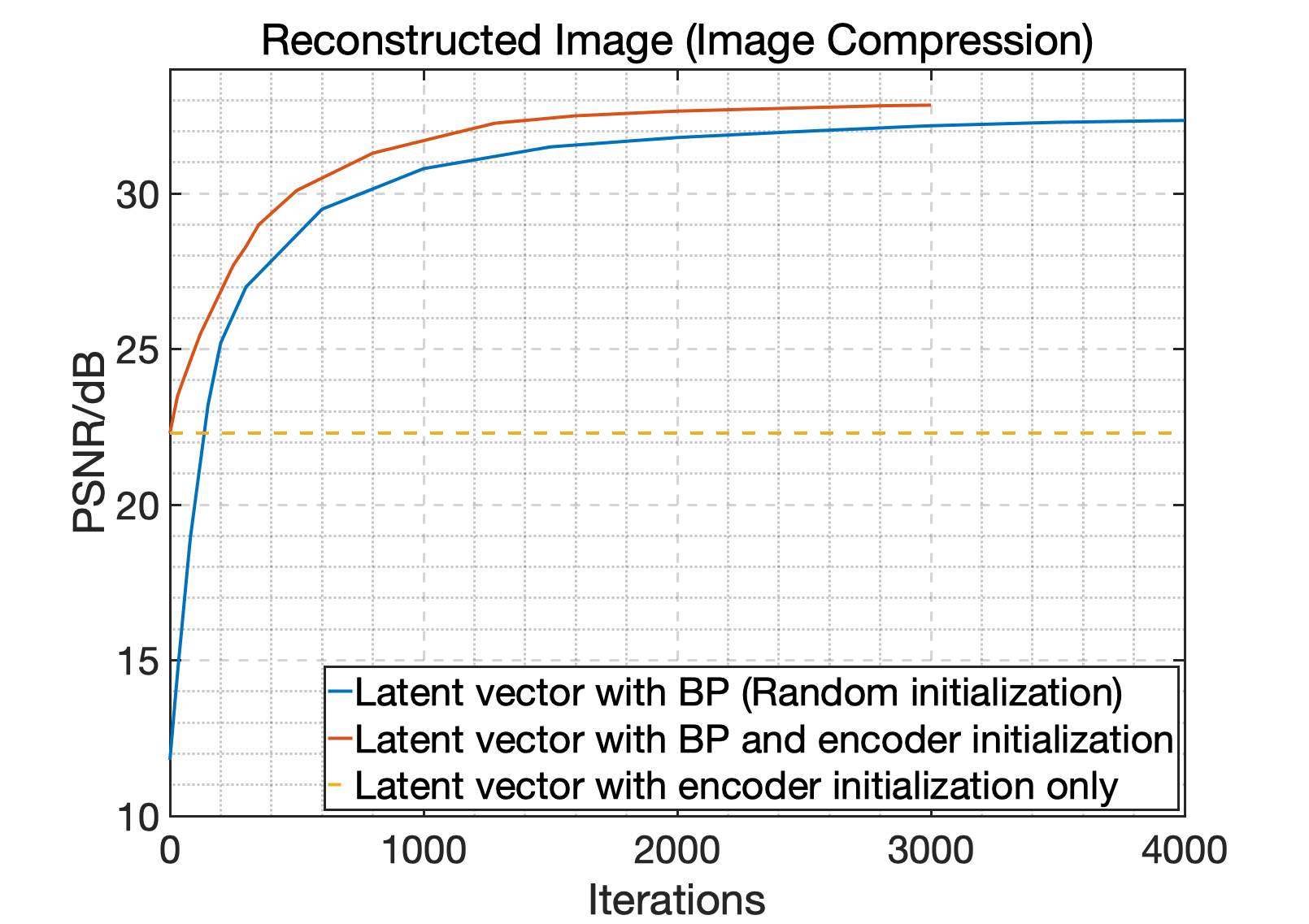}
    \end{subfigure}%
    \begin{subfigure}{.5\textwidth}
    \centering
    \includegraphics[width=\linewidth]{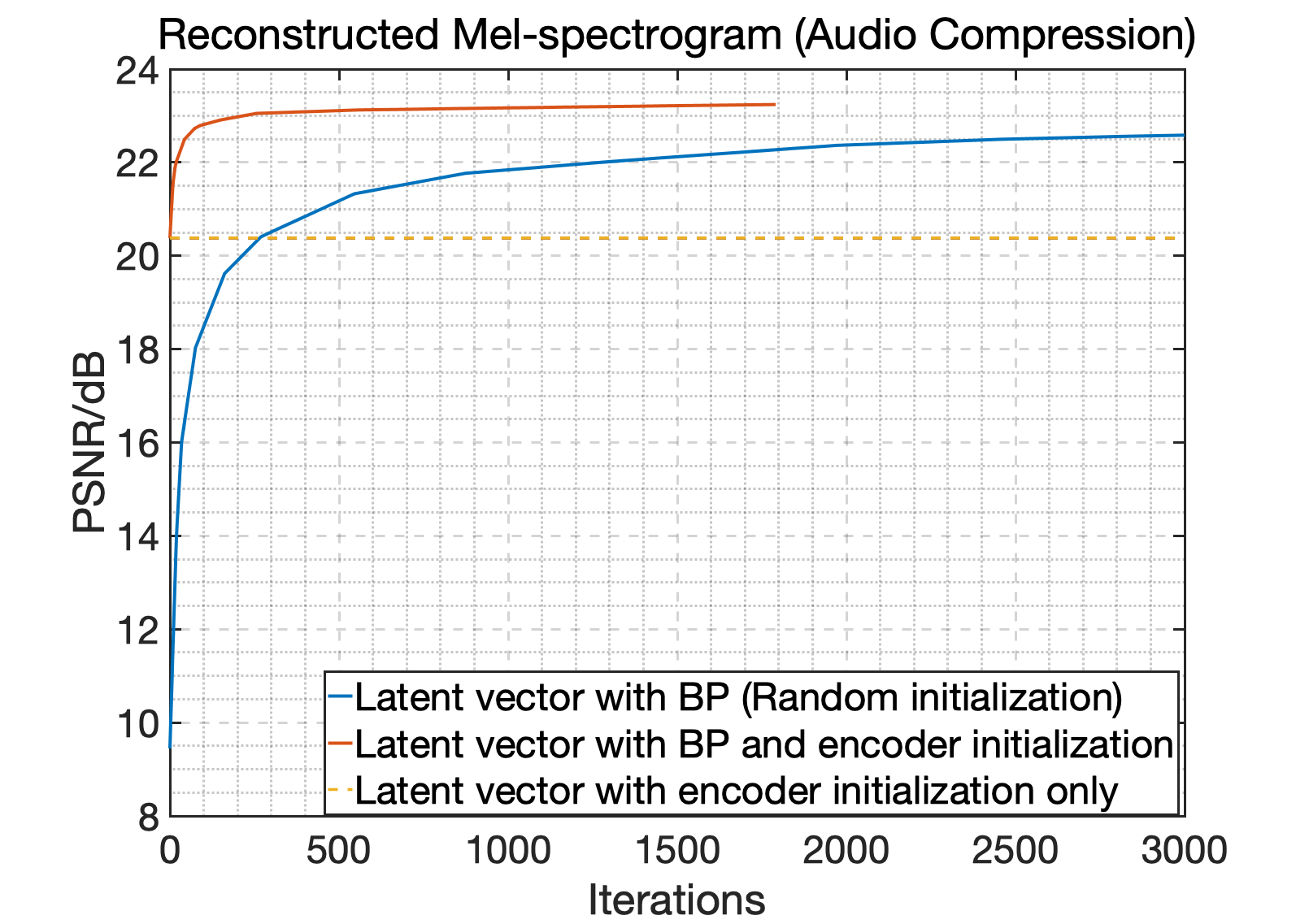}
    \end{subfigure}
\caption{Signal quality measured by PSNR vs. number of back-propagation iterations. Left: image compression, right: speech compression. Initializing latent vectors with an encoder before back-propagation (BP) can effectively reduce the number of iterations needed to find a good latent representation.}
\label{encoder}
\end{figure*}

\subsubsection{Evaluation Metrics}
In order to assess our speech compression quality, we use both subjective and objective metrics, including PSEQ, human evaluation under the guidance of MUSHRA, and phoneme recognition tests.

\textbf{PESQ}
 We evaluate the perceptual speech quality using PESQ \cite{rix2001perceptual}, which is an objective metric designed to predict the mean opinion score (MOS) for speech quality by an algorithm. It is adopted by ITU-T as a recommended standard metric. PESQ values range from -0.5 to 4.5, of which larger is better. PESQ is proposed for predicting the perceptual quality with acceptable accuracy for waveform-based compression and CELP codec for bit rates larger than 4 kilo-bits per second (kbps). 

\textbf{Subjective Evaluation}
We quantify the subjective quality of the (de)compressed speech with an experiment involving 20 users guided by Multiple Stimuli with Hidden Reference and Anchor (MUSHRA) \cite{vincent2006preliminary}. The users are asked to listen to 5 sets of audio clips. Each set of clips is presented with a labeled reference audio, a set of test samples, a hidden version of the reference, and one anchor. Based on their perceptual evaluation, the users are asked to score these test samples from 0 to 100. 

\textbf{Phoneme Recognition Evaluation}
  In addition, we perform phoneme recognition tests to measure the phoneme error rate (PER, lower the better). Phoneme recognition was performed on (de)compressed speech signal using the combination of SGMM and Dans deep neural networks in Kaldi \cite{povey2011kaldi}, and also using MLP and LSTM networks \cite{ravanelli2019pytorch}. These network network models take MFCC as the input and are trained with raw (uncompressed) audio data.

\subsubsection{Results and Analysis}
The speech compression evaluation results are summarized in Table \ref{audio_comp_perf} and Fig. \ref{audio_visual_2}. In this experiment, the original speech is sampled at 16k samples per second with 16-bit per sample, thus the original bit rate is 256 kbps. The compressed speech rate of 2 kbps is achieved by using a latent vector $\vz$ size of 512 and 16 non-uniform quantization levels for each element. While the proposed BPGAN based compression provides the lowest data rate of 2 kbps, it exhibits a better MUSHRA subjective quality score than the other methods with higher data rates except for Opus at 9 kbps (4.5$\times$ higher data rate than ours). We have observed that while the PESQ scores are similar among multiple methods, they do not accurately predict the (subjective) quality of the speech when the bit rate is low. The PER measured by phoneme recognition tests indicates that the error rate for the proposed BPGAN is superior (less PER degradation) to other compression methods while providing the lowest data rate.  

The experimental results in Table \ref{audio_comp_perf} show that our compression method significantly outperforms traditional codecs in both objective and subjective evaluations. For phoneme recognition tasks, our method demonstrates a clear advantage for all inference models even with at least $2\times$ smaller bit rate. This is because we compress audios in the spectral domain, preserving more context information with optimized feature loss through BP unlike other methods working in the time domain without an explicit objective function. 

\subsection{Ablation Studies}
\subsubsection{Quantization Analysis}
In Figure \ref{admm_loss},  we show the trade-off space between the loss of data quality (x-axis, lower is better) and the achievable bitrate (y-axis, without Huffman coding, lower is better). The gain of ADMM based non-uniform quantization is shown in the same figure. In all investigated cases, ADMM quantization manages to reduce the quality loss compared to direct quantization. One can notice that along the pareto-optimal line, the combination of the optimal latent vector dimension and the number of quantization levels per element changes for different bitrate targets.

In Fig. \ref{opt_comparison}, we compare the optimization loss (eq. (10)) vs. bitrate for direct K-means quantization, IHT optimization quantization, and our proposed ADMM optimization quantization. It demonstrates that the proposed ADMM method significantly (up to 2$\times$ in bit rate) outperforms the other methods.

\subsubsection{Effect of Different Latent Optimization Objectives}
\begin{table}[t]
\caption{Speech compression results with various loss functions during optimal latent vector search back-propagation.}
\centering
\resizebox{\columnwidth}{!}{
\begin{tabular}{c|c|c|c|c}
\hline
\multicolumn{1}{c}{\bf Method}  
&\multicolumn{1}{c}{\bf PESQ} 
&\multicolumn{1}{c}{\bf Kaldi PER\%}
&\multicolumn{1}{c}{\bf MLP PER \%}
&\multicolumn{1}{c}{\bf LSTM PER \%}
\\ \hline 
Original                    & 4.50  &18.7 &18.6 &15.4  \\
Feature Loss              & 2.92  &21.0 &20.6 &18.1\\
MSE Loss                  & 3.29  &21.5 &21.7 &19.9\\
\bf Combined Loss          & \bf3.25  &\bf 20.9 &\bf 20.8 &\bf 18.6\\
\hline
\end{tabular}
}
\label{speech_loss_comparison}
\end{table}
For image compression, we investigate the impact of using the combined objective function that includes the quality measure provided by the discriminator compared to the MSE only objective function. The results are visualized in Fig.\ref{image_visual_3}. It is evident that the MSE-only objective often induces blurry effect in the decompressed images leading to the loss of details. When incorporating the perceptual objective provided by the discriminator in addition to MSE, the iterative BP optimization process can find better latent representations that capture more accurate features and thus yield more faithful and visually pleasing reconstructions.

To study the influence of speech feature loss during the compression process, we performed ablation experiments on different back-propagation objectives (eq.(10)). The results are shown in Table \ref{speech_loss_comparison} where we compare the compressed speech quality/performance in PESQ and PER for the same 2 kbps rate  with feature-loss-only, MSE-loss-only, and a combination of MSE and feature losses as in eq.(10). Table \ref{speech_loss_comparison} shows that adding feature loss can significantly improve the performance on the speech recognition tasks compared with MSE-loss-only, and the combination loss achieves a good trade off between PESQ and phoneme recognition error rate. It is worth noting that the feature loss is calculated using  VGG-BLSTM neural network \cite{liu2019adversarial} that does not share the same structure in the phoneme recognition task evaluation networks, implying that the gain from the feature loss is not from overfitting. 

\subsubsection{Latent Initialization with Trained Encoder}
We evaluate the effectiveness of using an encoder for latent initialization by showing the compressed signal quality vs. the number of back-propagation iterations during image/speech compression tasks in Fig. \ref{encoder}. Compared with cases where the latent code are randomly initialized (in blue), encoder initialization (in red) improves the quality of compressed signals much faster to the saturated quality level, which is higher than that achieved from random initialization. The same figure also illustrates that the proposed iterative back-propagation brings significant gain compared to the one-shot encoder output (in yellow).

\section{Conclusion}
\label{conclusion}
We present BPGAN, a GAN-based unified signal compression method that produces compressed data with high fidelity at low bitrate by searching the optimal latent vector through iterative back-propagation. To improve the compression ratio, the proposed method applies ADMM with non-uniform quantization to search an optimal latent representation of the signal. The proposed BPGAN method first trains the generator network model in a GAN setup, then it iteratively updates and discretizes the optimal latent code through the pre-trained generator for each signal input for compression. Experiment results demonstrate that BPGAN-compressed signal exhibits significantly lower data rate and/or better signal quality compared to other methods evaluated with various metrics including neural network based image classification and speech phoneme recognition.



\ifCLASSOPTIONcompsoc
\else
\fi

\ifCLASSOPTIONcaptionsoff
  \newpage
\fi



%


\bibliographystyle{IEEEtran}
\bibliography{refs}
%
\begin{IEEEbiography}
[{\includegraphics[width=1in,height=1.25in,clip,keepaspectratio]{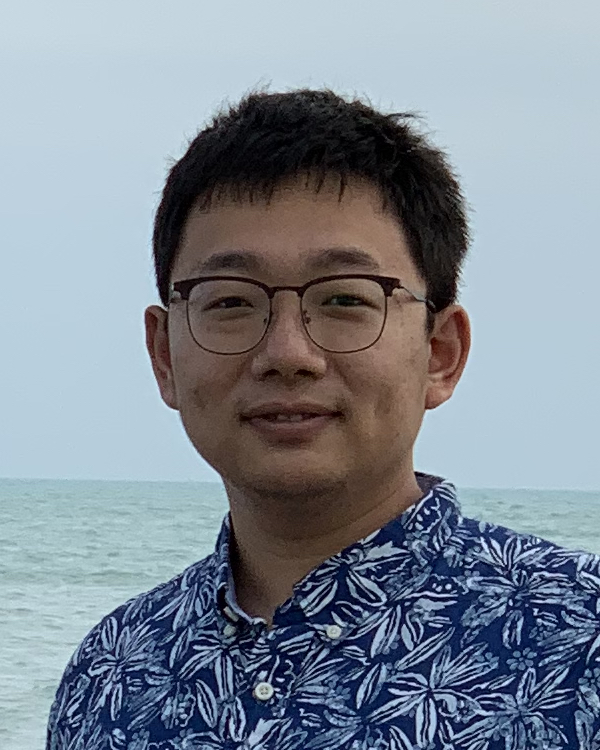}}]
{Bowen Liu} (Student Member, IEEE) received the M.S. degree in electrical engineering and computer science from the University of
Michigan, Ann Arbor, MI, USA, in 2018, where he is currently pursuing the Ph.D. degree with the Department of Electrical Engineering and Computer Science (EECS).

His research interests include deep learning, computer vision, signal processing, and their applications in low-power systems.
\end{IEEEbiography}

\begin{IEEEbiography}
[{\includegraphics[width=1in,height=1.25in,clip,keepaspectratio]{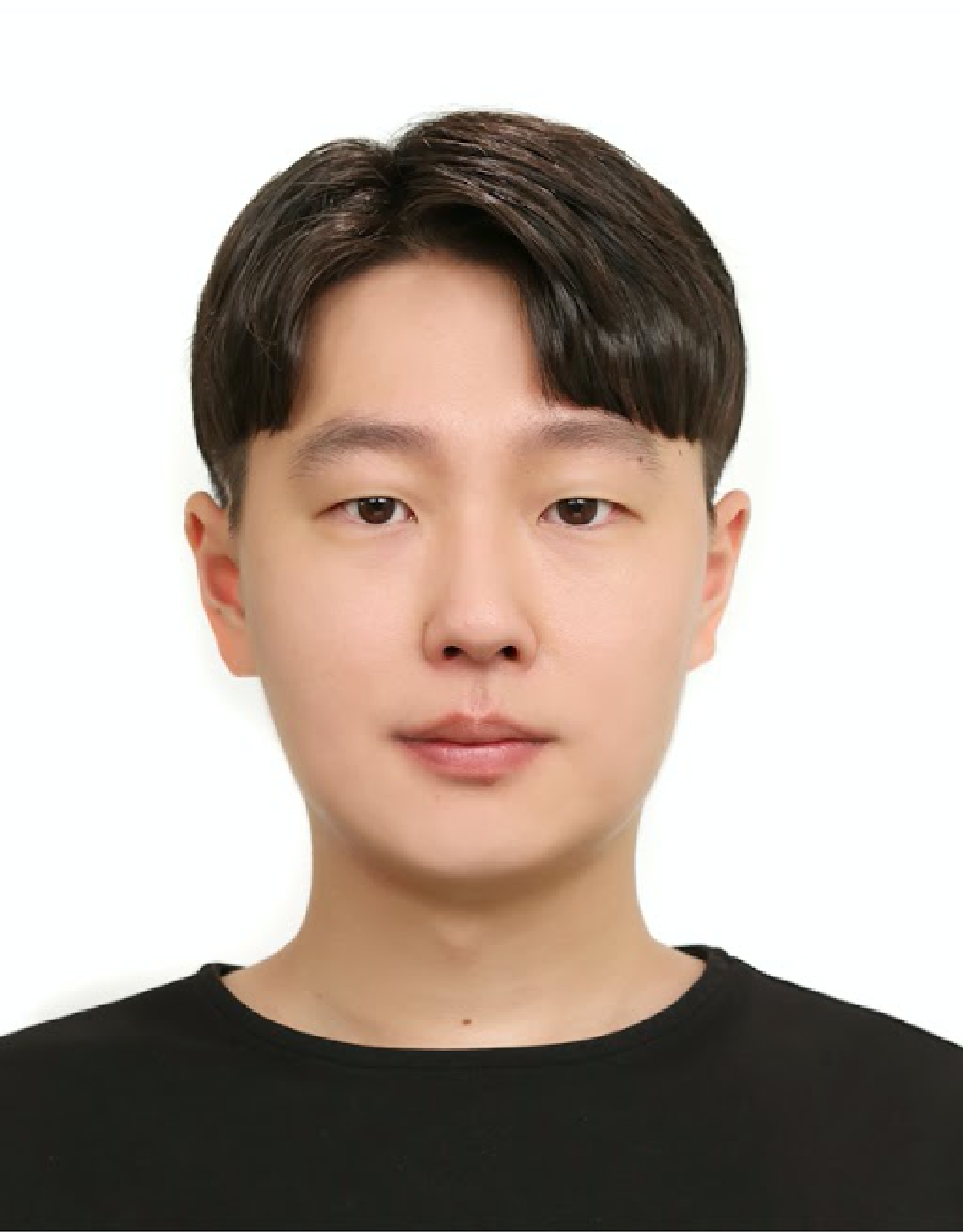}}]
{Changwoo Lee} received the B.S. and M.S. degree in electronic engineering from Hanyang University, Seoul, South Korea. He is working toward a Ph.D. degree in electrical engineering and computer science at the University of Michigan, Ann Arbor, MI, USA. His research interests include machine learning and signal processing.
\end{IEEEbiography}

\begin{IEEEbiography}
[{\includegraphics[width=1in,height=1.25in,clip,keepaspectratio]{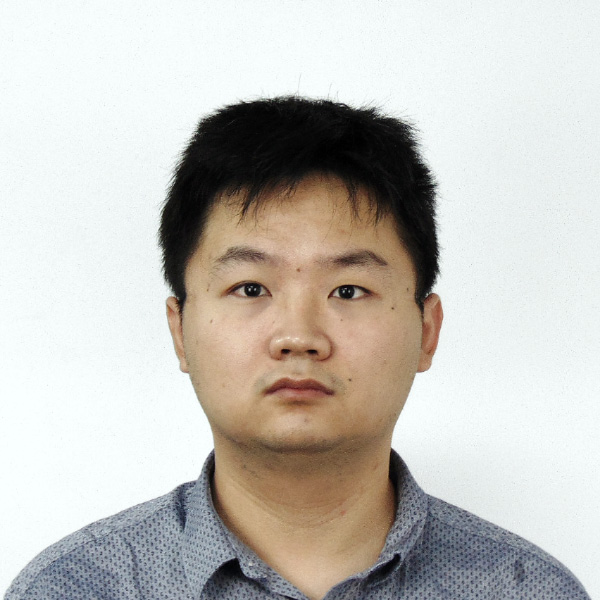}}]
{Ang Cao} received the B.S. degree in Electrical Engineering from the Wuhan University, Wuhan, China, in 2018 and M.S. degree in Electrical and   Computer Engineering from the University of Michigan, Ann Arbor, MI, in 2020. He is currently a Ph.D. student of Computer Science and Engineering at the University of Michigan, Ann Arbor, MI.
His current research interests include computer vision, signal processing, and inverse problems. 
\end{IEEEbiography}

\begin{IEEEbiography}
[{\includegraphics[width=1in,height=1.25in,clip,keepaspectratio]{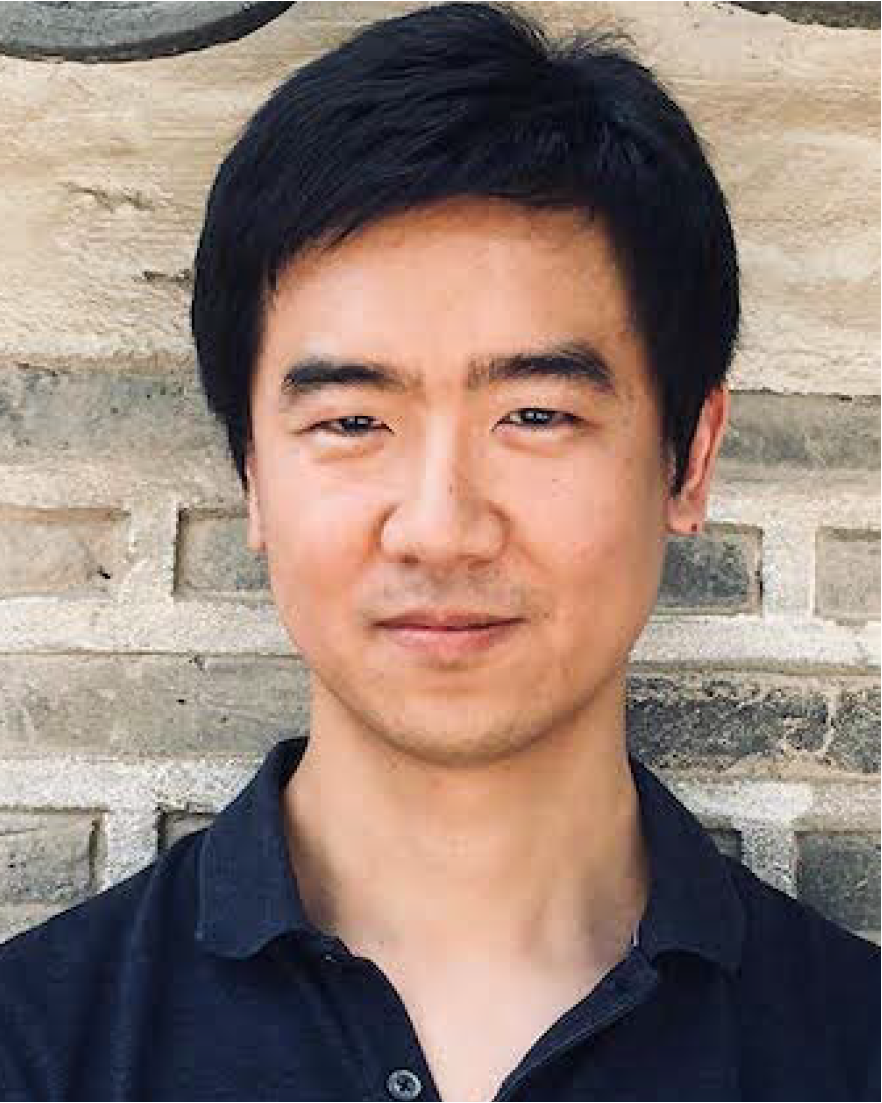}}]
{HunSeok Kim} (Member, IEEE) received the B.S. degree in electrical engineering from Seoul National  University, Seoul, South Korea, in 2001, and the Ph.D. degree in electrical engineering from the University of California at Los Angeles (UCLA), Los Angeles, CA, USA, in 2010.

He is currently an Assistant Professor with the University of Michigan, Ann Arbor, MI, USA. His research focuses on system analysis, novel algorithms, and very-large-scale integration (VLSI) architectures for low-power/high-performance wireless communications, signal processing, computer vision, and machine learning systems.

Dr. Kim was a recipient of the 2018 Defense Advanced Research Projects Agency (DARPA) Young Faculty Award (YFA) and the National Science Foundation (NSF) Faculty Early Career Development (CAREER) Award 2019. He is an Associate Editor of the IEEE TRANSACTIONS ON MOBILE COMPUTING, the IEEE TRANSACTIONS ON GREEN COMMUNICATIONS AND NETWORKING, and the IEEE SOLID STATE CIRCUITS LETTERS.
\end{IEEEbiography}

\end{document}

%% file: math_commands.tex
\usepackage{amsmath,amsfonts,bm}

\def\eqref#1{equation~\ref{#1}}

\def\1{\bm{1}}

\def\va{{\bm{a}}}

\def\ve{{\bm{e}}}

\def\vu{{\bm{u}}}

\def\vx{{\bm{x}}}
\def\vy{{\bm{y}}}
\def\vz{{\bm{z}}}

\def\mS{{\bm{S}}}

\DeclareMathAlphabet{\mathsfit}{\encodingdefault}{\sfdefault}{m}{sl}
\SetMathAlphabet{\mathsfit}{bold}{\encodingdefault}{\sfdefault}{bx}{n}

\newcommand{\E}{\mathbb{E}}

\newcommand{\R}{\mathbb{R}}

